\documentclass[a4paper,fleqn,usenatbib]{mnras}

\usepackage{newtxtext,newtxmath}
\usepackage[T1]{fontenc}
\usepackage{ae,aecompl}

\usepackage{graphicx}	% Including figure files
\usepackage{amsmath}	% Advanced maths commands
\usepackage{amssymb}	% Extra maths symbols

\newcommand{\msun}{M$_\odot$}
\newcommand{\rsun}{R$_\odot$}
\newcommand{\lsun}{L$_\odot$}

\newcommand{\mearth}{M$_\oplus$}
\newcommand{\rearth}{R$_\oplus$}
\newcommand{\kms}{\ensuremath{\rm km\,s^{-1}}}
\newcommand{\ms}{\ensuremath{\rm m\,s^{-1}}}

\newcommand{\SMW}{\ensuremath{S _{\rm MW}}}
\newcommand{\rhk}{\ensuremath{\log R'_{\rm HK}}}

\newcommand{\thisstar}{K2-263}
\newcommand{\thisplanet}{K2-263~b}

\title[\thisplanet]{\thisplanet: A 50-day period sub-Neptune with a mass measurement using HARPS-N}

\author[A. Mortier et al.]{
A.\,Mortier$^{1}$\thanks{E-mail: am352@st-andrews.ac.uk},
A.\,S.\,Bonomo$^{2}$,
V.\,M.\,Rajpaul$^{3}$,
L.\,A.\,Buchhave$^{4}$,
A.\,Vanderburg$^{5,\dagger}$,
\newauthor
L.\,Zeng$^{6}$,
M.\,L\'{o}pez-Morales$^{7}$,
L.\,Malavolta$^{8,9}$,
A.\,Collier Cameron$^{1}$,
C.\,D.\,Dressing$^{10}$,
\newauthor
P.\,Figueira$^{11,12}$,
V.\,Nascimbeni$^{9,8}$,
K.\,Rice$^{13,14}$,
A.\,Sozzetti$^{2}$,
C.\,Watson$^{15}$,
\newauthor
L.\,Affer$^{16}$,
F.\,Bouchy$^{17}$,
D.\,Charbonneau$^{7}$,
A.\,Harutyunyan$^{18}$,
R.\,D.\,Haywood$^{7,\dagger}$,
\newauthor
J.\,A.\,Johnson$^{7}$,
D.\,W.\,Latham$^{7}$,
C.\,Lovis$^{17}$,
A.\,F.\,Martinez Fiorenzano$^{18}$,
M.\,Mayor$^{17}$,
\newauthor
G.\,Micela$^{16}$,
E.\,Molinari$^{19}$,
F.\,Motalebi$^{17}$,
F.\,Pepe$^{17}$,
G.\,Piotto$^{9,8}$,
D.\,Phillips$^{7}$,
\newauthor
E.\,Poretti$^{18,20}$,
D.\,Sasselov$^{7}$,
D.\,S\'egransan$^{17}$,
S.\,Udry$^{17}$,
\\
$^{1}$Centre for Exoplanet Science, SUPA, School of Physics and Astronomy, University of St Andrews, St Andrews KY16 9SS, UK\\
$^{2}$INAF - Osservatorio Astrofisico di Torino, via Osservatorio 20, 10025 Pino Torinese, Italy\\
$^{3}$Astrophysics Group, Cavendish Laboratory, University of Cambridge, J.\.J.\,Thomson Avenue, Cambridge CB3 0HE, UK\\
$^{4}$DTU Space, National Space Institute, Technical University of Denmark, Elektrovej 328, DK-2800 Kgs. Lyngby, Denmark\\
$^{5}$Department of Astronomy, The University of Texas at Austin, 2515 Speedway, Stop C1400, Austin, TX 78712, USA\\
$^{6}$Department of Earth and Planetary Sciences, Harvard University, Cambridge, MA, 02138, USA\\
$^{7}$Harvard-Smithsonian Center for Astrophysics, 60 Garden Street, Cambridge, MA 01238, USA\\
$^{8}$INAF - Osservatorio Astronomico di Padova, Vicolo dell'Osservatorio 5, 35122 Padova, Italy\\
$^{9}$Dipartimento di Fisica e Astronomia ``Galileo Galilei", Universita' di Padova, Vicolo dell'Osservatorio 3, I-35122 Padova, Italy\\
$^{10}$Astronomy Department, University of California, Berkeley, CA 94720, USA\\
$^{11}$European Southern Observatory, Alonso de Cordova 3107, Vitacura,Santiago, Chile\\
$^{12}$Instituto de Astrofisica e Ciencias do Espa\c co, CAUP, Universidade do Porto, Rua das Estrelas, PT4150-762 Porto, Portugal\\
$^{13}$SUPA, Institute for Astronomy, Royal Observatory, University of Edinburgh, Blackford Hill, Edinburgh EH93HJ, UK\\
$^{14}$Centre for Exoplanet Science,  University of Edinburgh,  Edinburgh,  UK\\
$^{15}$Astrophysics Research Centre, School of Mathematics and Physics, Queen's University Belfast, Belfast, BT7 1NN, UK\\
$^{16}$INAF - Osservatorio Astronomico di Palermo, Piazza del Parlamento 1, 90134 Palermo, Italy\\
$^{17}$Observatoire Astronomique de l'Universit\'e de Gen\`eve, Chemin des Maillettes 51, Sauverny, CH-1290, Switzerland\\
$^{18}$INAF - Fundaci\'on Galileo Galilei, Rambla Jos\'e Ana Fernandez P\'erez 7, E-38712 Bre\~na Baja, Tenerife, Spain\\
$^{19}$INAF - Osservatorio Astronomico di Cagliari, via della Scienza 5, 09047, Selargius, Italy\\
$^{20}$INAF - Osservatorio Astronomico di Brera, Via E. Bianchi 46, 23807 Merate (LC), Italy\\
$\dagger$NASA Sagan Fellow
}

\date{Accepted 2018 August 24. Received 2018 August 24; in original form 2018 August 10}

\pubyear{2018}

\begin{document}
\label{firstpage}
\pagerange{\pageref{firstpage}--\pageref{lastpage}}
\maketitle

\begin{abstract}
This paper reports on the validation and mass measurement of \thisplanet, a sub-Neptune orbiting a quiet G9V star. Using {\it K2} data from campaigns C5 and C16, we find this planet to have a period of $50.818947\pm 0.000094$\,days and a radius of $2.41\pm0.12$\,\rearth. We followed this system with HARPS-N to obtain 67 precise radial velocities. A combined fit of the transit and radial velocity data reveals that \thisplanet\ has a mass of $14.8\pm3.1$\,\mearth. Its bulk density ($5.7_{-1.4}^{+1.6}$\,g\;cm$^{-3}$) implies that this planet has a significant envelope of water or other volatiles around a rocky core. \thisplanet\ likely formed in a similar way as the cores of the four giant planets in our own Solar System, but for some reason, did not accrete much gas. The planetary mass was confirmed by an independent Gaussian process-based fit to both the radial velocities and the spectroscopic activity indicators. \thisplanet\ belongs to only a handful of confirmed {\it K2} exoplanets with periods longer than 40 days. It is among the longest periods for a small planet with a precisely determined mass using radial velocities.
\end{abstract}

\begin{keywords}
planets and satellites: \thisplanet\ -- techniques: photometric -- techniques: radial velocities -- techniques: spectroscopic
\end{keywords}

%%%%%%%%%%%%%%%%%%%%%%%%%%%%%%%%%%%%%%%%%%%%%%%%%%

%%%%%%%%%%%%%%%%% BODY OF PAPER %%%%%%%%%%%%%%%%%%

\section{Introduction}\label{intro}

Both the {\it Kepler} mission and its revived version, the {\it K2} mission, have discovered thousands of exoplanets, uncovering an exciting diversity in the exoplanet population \citep[e.g. ][]{Morton16,Mayo18}. The modified {\it K2} mission differs from the original {\it Kepler} mission in that it does not stare at the same field, but instead visits multiple fields in the Ecliptic Plane, each for about 80 days. This limited timespan makes the mission sensitive to short-period planets only. 

Only a handful of K2 exoplanets with periods longer than 40 days (half the timespan of a \emph{K2} campaign) have been reported and validated\footnote{according to \url{http://archive.stsci.edu/k2/published_planets/}}. The planet with the longest period within the K2 campaign timespan is K2-118\,b. It has a period of $50.921$\,days and a radius of $2.49$\,\rearth \citep{Dre17}. The faintness of the star ($V\sim14$) impedes obtaining precise radial velocities (RVs). The other validated long-period exoplanets from K2 are the three outer planets (each showing a monotransit) in the five-planet system orbiting HIP41378 \citep{Vand16b} with estimated periods of $156, 131, 324$\,days for planets d, e, and f, respectively. No mass measurements have been reported on this system yet.

Precise and accurate masses for planets similar to Earth in size with a variety of orbital periods are essential to understand the transition between rocky and non-rocky planets for small planets. Recently, a gap was found around 2\,\rearth in the distribution of planetary radii of Kepler planets \citep[e.g. ][]{Ful17,Zeng17b,VanE18,Ful18}. Planets with radii below that gap are most likely rocky or Earth-like in composition. However, without a value for the planetary mass, the composition of the planets above the gap remains uncertain.

Having a well-characterised sample of small planets spanning a broad variety of parameters, such as orbital period, planetary mass, planetary radius, and various stellar parameters (mass, radius, chemical abundances, ...), can shed light on the formation and evolution history of these planets. This can include their formation location (in terms of the snow line), the amount of planetary migration, and the effects of photo-evaporation amongst other scenarios.

In this paper we report on a four-sigma mass measurement of \thisplanet. This planet was labeled as a small planetary candidate in \citet{Mayo18} with an orbital period of $50.82$\,days and a preliminary planetary radius above the radius gap.

This paper is structured as follows. Section \ref{observations} describes the obtained data, both from photometry and spectroscopy. We validate the transit in Section \ref{valid}. Stellar properties, including stellar activity indicators, are discussed in Section \ref{star}. Sections \ref{combo} and \ref{rv} describe the two analyses we performed on the light curve and RVs. Finally, we discuss and conclude in Section \ref{concl}.

\section{Observations}\label{observations}

We recovered photometric observations from K2 and obtained spectroscopic observations from HARPS-N for \thisstar.

\subsection{K2 Photometry} \label{k2}

\thisstar\ was observed on two occasions with NASA's {\it K2} mission. During Campaign 5 from 2015 April 27 till 2015 July 10, it was observed in long cadence mode ($29.4$\,min) only\footnote{Guest Observer programmes: GO5007\_LC, GO5029\_LC, GO5033\_LC, GO5104\_LC, GO5106\_LC, and GO5060\_LC.}. Campaign 16 (from 2017 December 7 till 2018 Feb 25) observed \thisstar\ both in long cadence and in short cadence mode ($1$\,min)\footnote{Guest Observer programmes: GO16009\_LC, GO16011\_LC, GO16015\_LC, GO16020\_LC, GO16021\_LC, GO16101\_LC, GO16009\_SC, GO16015\_SC, and GO16101\_SC}.

The data were obtained via the Mikulski Archive for Space Telescopes (MAST\footnote{\url{https://archive.stsci.edu/k2/}}) and subsequently processed following the procedures described in \citet{Vand14} and \citet{Vand16a}. In short, we initially produced a first-pass light curve. Upon a periodicity search, a transit signal was recovered with a periodicity of $50.8$\, days.  We then used this rough solution as a basis to extract the final light curve where we simultaneously fitted the long-term instrumental trends, the 6 hour thruster systematics, and the transits\footnote{The full light curves can be obtained from \url{https://www.cfa.harvard.edu/~avanderb/k2c5/ep211682544.html} and \url{https://www.cfa.harvard.edu/~avanderb/k2c16/ep211682544.html}}. 

Due to the limited observing period of 80\,days for each {\it K2} Campaign, only two transit events occur per Campaign. The second transit in Campaign 16 could not be extracted reliably due to a brief jump in the spacecraft pointing jitter. Consequently, we have only 3 transit events for this target, two with long cadence and one with both long and short cadence, as seen in Figure \ref{fig_transitfit}. 

\subsection{HARPS-N Spectroscopy} \label{harps}

\begin{table*}
\caption{Sample of measured radial velocities and activity indicators for \thisstar. The full table is available online.}            
\label{tab_rv}
\begin{tabular}{lllllllll}        
\hline\hline
Time & RV & $\sigma_{\text{RV}}$ & FWHM & BIS & \SMW & $\sigma_{\text{S}}$ & $\log{R'_{HK}}$ & $\sigma_{\text{RHK}}$ \\
{[}BJD] & [\kms] & [\kms] & [\kms] & [\kms] & [dex] & [dex] & [dex] & [dex]\\
\hline
$2457379.631593$ & $29.99716$ & $0.00176$ & $6.10295$ & $-0.00840$ & $0.160672$ & $0.004935$ & $-5.019180$ & $0.026653$ \\
$2457380.645277$ & $29.99650$ & $0.00267$ & $6.11709$ & $-0.00702$ & $0.160894$ & $0.009688$ & $-5.017983$ & $0.052179$ \\
$2457381.651863$ & $29.99933$ & $0.00262$ & $6.09290$ & $-0.00550$ & $0.157909$ & $0.009437$ & $-5.034365$ & $0.052781$ \\
$2457382.681749$ & $30.00122$ & $0.00401$ & $6.05141$ & $-0.00101$ & $0.165530$ & $0.016205$ & $-4.993705$ & $0.082533$ \\
$2457385.645590$ & $29.99342$ & $0.00320$ & $6.09260$ & $-0.00135$ & $0.150686$ & $0.012838$ & $-5.076767$ & $0.079166$ \\
\ldots \\
\hline\hline
\end{tabular}
\end{table*}

We obtained 67 spectra of the G9V host with HARPS-N \citep{Cos12}, installed at the Telescopio Nazionale Galileo (TNG) in La Palma, Spain. The spectra were taken between December 2015 and January 2018, each with an exposure time of 30 minutes. The spectra have a mean signal-to-noise ratio of 37 in order 50 (centered around 5650\,\AA). RVs were determined with the dedicated pipeline, the Data Reduction Software \citep[DRS - ][]{Bara96} where a G2 mask was used to calculate the weighted cross correlation function \citep[CCF - ][]{Pepe02b}. The RV errors are photon-limited with an average RV error of $2.8$\,\ms whilst the RMS of the RVs is $3.9$\,\ms.

The DRS also provides some activity indicators, such as the full width at half maximum (FWHM) of the CCF, the CCF line bisector inverse slope (BIS), the CCF contrast, and the Mount Wilson S-index (\SMW) and chromospheric activity indicator \rhk from the \ion{Ca}{ii} H\&K lines \citep[see e.g. ][]{Noy84b,Que01,Que09}. Error values for the FWHM, BIS, and contrast were calculated following the recommendations of \citet{Sant15}.

We checked each HARPS-N observation for moonlight contamination with a procedure outlined in \citet{Mal17}. Following this, we decided to discard the last 4 points, taken in January 2018 during a near-full Moon. 

\begin{figure}
\begin{center}
\includegraphics[width=\linewidth]{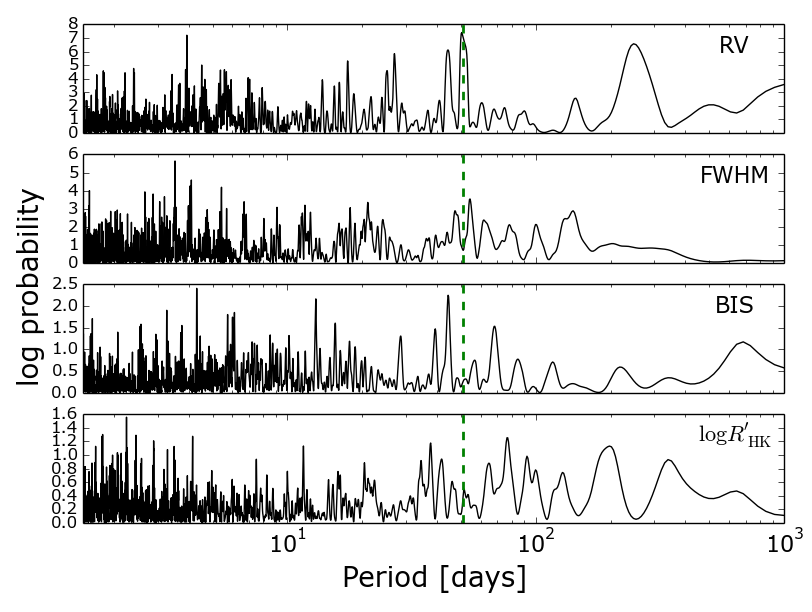}
\caption[]{Top to bottom: BGLS periodogram of the time series of RV, FWHM, BIS, \rhk. The green dashed vertical line indicates the orbital period of the transiting planet.}
\label{fig_bgls}
\end{center}
\end{figure}

All RVs with their errors and activity indicators are listed in Table \ref{tab_rv}. We computed a Bayesian Generalized Lomb-Scargle periodogram \citep[BGLS - ][]{ME15} of the data, as shown in the top plot of Figure \ref{fig_bgls}. The transit period of $50.8$\,days is also found to be the strongest periodicity in the RV data.

\section{Transiting planet validation}\label{valid}

In the recent work of \citet{Mayo18}, \thisplanet\ was found to be a planet candidate with a False Positive Probability of $0.00292$ using the probabilistic algorithm \textsc{vespa} \citep{Morton12}. They used a threshold $0.001$ to validate the transit signal to be attributed to an exoplanet. Their identified planet period is $50.819 \pm 0.002$\,days with a mid-transit time of $2457145.568$\,BJD. The full outcome of their analysis can be found in \citet{Mayo18b}.

The dominant false positive scenario that remained is that the star is an eclipsing binary. However, our HARPS-N observations conclusively rule that out. If the transit signal were due to an eclipsing binary, we would expect large (on the order of several \kms) RV variations. With an RV RMS of only $3.9$\,\ms, we can eliminate the scenario of an eclipsing binary. By including this eliminated scenario in the results of \citet{Mayo18b}, the false positive probability decreases to $0.000303$, less than 1 in 1000, and thus statistically validating the presence of a planet orbiting \thisstar.

\section{Stellar properties}\label{star}

\thisstar\ is a G9 dwarf star, with an apparent V magnitude of $11.61$. The star is located at a distance of $163\pm1$\,pc as obtained via the new and precise parallax from the Gaia mission second data release \citep{Gaia16,Gaia2}. All stellar properties are listed in Table \ref{tab_star}.

\begin{table}
%\vspace{-0.4cm}
\caption{\thisstar\ stellar properties}            
\label{tab_star}
\begin{tabular}{l l l}        
\hline\hline
Parameter & Value & Source \\
\hline
\multicolumn{3}{l}{\emph{Designations and coordinates}} \\
EPIC ID & 211682544 & EPIC \\
\emph{K2} ID & 263 & \\
2-MASS ID & J08384378+1540503 & \\
RA (J2000)   & 08:38:43.78 & 2MASS\\
Dec (J2000) &  15:40:50.4 &  2MASS\\
\hline
\multicolumn{3}{l}{\emph{Magnitudes and parallax}}\\
B & $12.35 \pm 0.03$ & APASS\\
V & $11.61 \pm 0.04$ & APASS\\
\emph{Kepler} magnitude & $11.41$ & EPIC \\
J & $10.22 \pm 0.02$ & 2MASS\\
H & $9.81 \pm 0.02$ & 2MASS\\
K & $9.75 \pm 0.02$ & 2MASS\\
Parallax $\pi$ & $6.1262 \pm 0.0514$ & Gaia DR2\\
Distance $d$ [pc] & $163.2 \pm 1.4$ & 1 \\
\hline
\multicolumn{3}{l}{\emph{Atmospheric parameters}: Effective temperature $T_{\rm{eff}}$,} \\
\multicolumn{3}{l}{surface gravity log\,$g$, metallicity $[\rm{Fe/H}]$, projected} \\
\multicolumn{3}{l}{rotational velocity v$\sin$\,i, microturbulence $\xi_t$} \\
$T_{\rm{eff}}$ [K] & $5372 \pm 73$ & 2 \\
log\,$g$ [cgs] &  $4.58 \pm 0.13$ & 2 \\
$[\rm{Fe/H}]$ [dex] & $-0.08  \pm 0.05$ & 2 \\
$\xi_t$ [km/s] & $0.76 \pm 0.08$ & 2 \\
$T_{\rm{eff}}$ [K] & $5365 \pm 50$ & 3 \\
log\,$g$ [cgs] &  $4.45 \pm 0.10$ & 3 \\
$[\rm{m/H}]$ [dex] & $-0.07 \pm 0.08$ & 3 \\
v$\sin$\,i [km/s] & $<2.0$ & 3 \\
\multicolumn{3}{l}{\emph{Adopted averaged parameters}} \\
$T_{\rm{eff}}$ [K] & $5368 \pm 44$ &  \\
log\,$g$ [cgs] &  $4.51 \pm 0.08$ &  \\
$[\rm{m/H}]$ [dex] & $-0.08 \pm 0.05$ &  \\
\hline
\multicolumn{3}{l}{\emph{Mass, radius, age, luminosity}} \\
M$_*$ [\msun] &  $0.86 \pm 0.03$ & 4 \\
R$_*$ [\rsun] & $0.84 \pm 0.02$ & 4  \\
Age $t$ [Gyr]  & $7\pm4$ & 4 \\
L$_*$ [\lsun] &  $0.55 \pm 0.02$ & 5 \\
R$_*$ [\rsun] & $0.86 \pm 0.02$ & 5  \\
M$_*$ [\msun] &  $0.87 - 0.89$ & 6 \\
\multicolumn{3}{l}{\emph{Adopted averaged parameters}} \\
M$_*$ [\msun] &  $0.88 \pm 0.03$ &  \\
R$_*$ [\rsun] & $0.85 \pm 0.02$ &  \\
$\rho_{*}$ [g\,cm$^{-3}$] & $2.02 \pm 0.16$ &  \\
\hline\hline
\end{tabular}
\newline \centering 1: Using the Gaia DR2 parallax, 2: ARES+MOOG \citep{Sou14}, with the surface gravity corrected following \citep{ME14}, 3: SPC \citep{Buch12,Buch14}, 4: Using PARSEC isochrones \citep{Das06,Bre12}, 5: Using distance, apparent magnitude, and bolometric correction, 6: Relations of \citet{Moya18}
\end{table}

\subsection{Atmospheric parameters} \label{star2}

We used two different methods to determine the stellar atmospheric parameters. The first method, explained in more detail in \citet{Sou14} and references therein, is based on equivalent widths. We added all HARPS-N spectra together for this method. We automatically determined the equivalent widths of a list of iron lines (\ion{Fe}{1} and \ion{Fe}{2}) \citep{Sou11a} using ARESv2 \citep{Sou15}. The atmospheric parameters were then determined via a minimisation procedure, using a grid of ATLAS plane-parallel model atmospheres \citep{Kur93} and the 2014 version of the MOOG code\footnote{\url{http://www.as.utexas.edu/~chris/moog.html}} \citep{Sne73}, assuming local thermodynamic equilibrium. The surface gravity was corrected based on the value for the effective temperature following the recipe explained in \citet{ME14}. We quadratically added systematic errors to our precision errors, intrinsic to our spectroscopic method. For the effective temperature we added a systematic error of $60$\,K, for the surface gravity $0.1$\,dex, and for metallicity $0.04$\,dex \citep{Sou11a}.

Additionally, we used the Stellar Parameter Classification tool \citep[SPC - ][]{Buch12,Buch14} to obtain the atmospheric parameters. SPC was run on 63 individual spectra after which the values were averaged, weighted by their signal-to-noise ratio. The results agree remarkably well with the values from the ARES+MOOG method. As SPC is a spectrum synthesis method, it also determined a rotational velocity. This showed that \thisstar\ is a slowly rotating star with $v\sin i<2$\,km/s.

We finally adopted the average of the parameters obtained with both methods for subsequent analyses in this work. \thisstar\ has a temperature of $5368\pm44$\,K, a metallicity of [m/H] $=-0.08\pm0.05$, and a surface gravity of $\log g = 4.51\pm0.08$ (cgs).

\subsection{Mass and radius} \label{star3}

We obtained values for the stellar mass and radius by fitting stellar isochrones, using the adopted atmospheric parameters from the previous section, the apparent V magnitude and the new and precise Gaia parallax. We used the PARSEC isochrones \citep{Bre12} and a Bayesian estimation method \citep{Das06} through their web interface\footnote{http://stev.oapd.inaf.it/cgi-bin/param\_1.3}. From this, we obtain a stellar mass of $0.86\pm0.03$\,\msun\ and a stellar radius of $0.84\pm 0.02$\,\rsun. Through this isochrone fitting, we also determined a stellar age of $7\pm4$\,Gyr. 

The very precise Gaia parallax allows for a direct calculation of the absolute magnitude of \thisstar. Extinction is negligible according to the dustmaps of \citet{Green18}. Stellar luminosity can then be calculated, for which we used the bolometric correction from \citet{Flo96} with corrected coefficients from \citet{Tor10b}. We get a stellar luminosity of $L_{\ast} = 0.55 \pm 0.02$\,\lsun. Combining with the effective temperature, this results in a stellar radius of $R_{\ast}=0.86 \pm0.02$\,\rsun.

We furthermore employed the relations by \citet{Moya18} to obtain a value for the stellar mass. We used three logarithmic relations between stellar mass and stellar luminosity, metallicity, and effective temperature. For these three cases, we obtained values for the stellar mass between $0.87$ and $0.89$\,\msun\ which are in agreement with the mass value calculated above.

For the remainder of this work, we adopted the average parameters of $R_{\ast}=0.85 \pm0.02$\,\rsun\ and $M_{\ast}=0.88 \pm0.03$\,\msun, for the radius and mass of the star.

\subsection{Stellar activity} \label{act}

\thisstar\ is a relatively quiet star as evident from the average mean value of \rhk \citep[$-5.00\pm0.05$ - see e.g. ][]{Mama08}. The light curve shows no periodic variations that can be used to estimate a rotation period for this star. However, we can use the average \rhk, together with the colour B-V to estimate the rotation period. The empirical relationships of \citet[][their Eqs. 3 and 4]{Noy84b} provide a rotation period of 35\,days whilst the recipe of \citet[][their Eq. 5]{Mama08} gives 37\,days. This is in agreement with the low rotational velocity determined from SPC in Section \ref{star2}.

We investigated the periodicities in the time series of the main activity indicators (FWHM, BIS, \rhk). Figure \ref{fig_bgls} shows the BGLS periodograms of all three indicators. There is some variability in the indicators, but no strong periodic signals, which agrees with this star being quiet. The planet period is furthermore not present in either of the indicators, giving us confidence that the 50\,d periodic signal in the RVs can indeed be attributed to the transiting planet.

Correlations between the RVs and activity indicators can be a sign of stellar variability in the RVs. We calculated the Spearman correlation coefficient for RV versus FWHM, BIS, and \rhk and find them to be $-0.09, -0.14$, and $0.04$, respectively, indicating no strong correlation with these indicators.

\section{Combined transit and RV analysis}\label{combo}

\begin{figure*}
\begin{center}
\includegraphics[width=\textwidth]{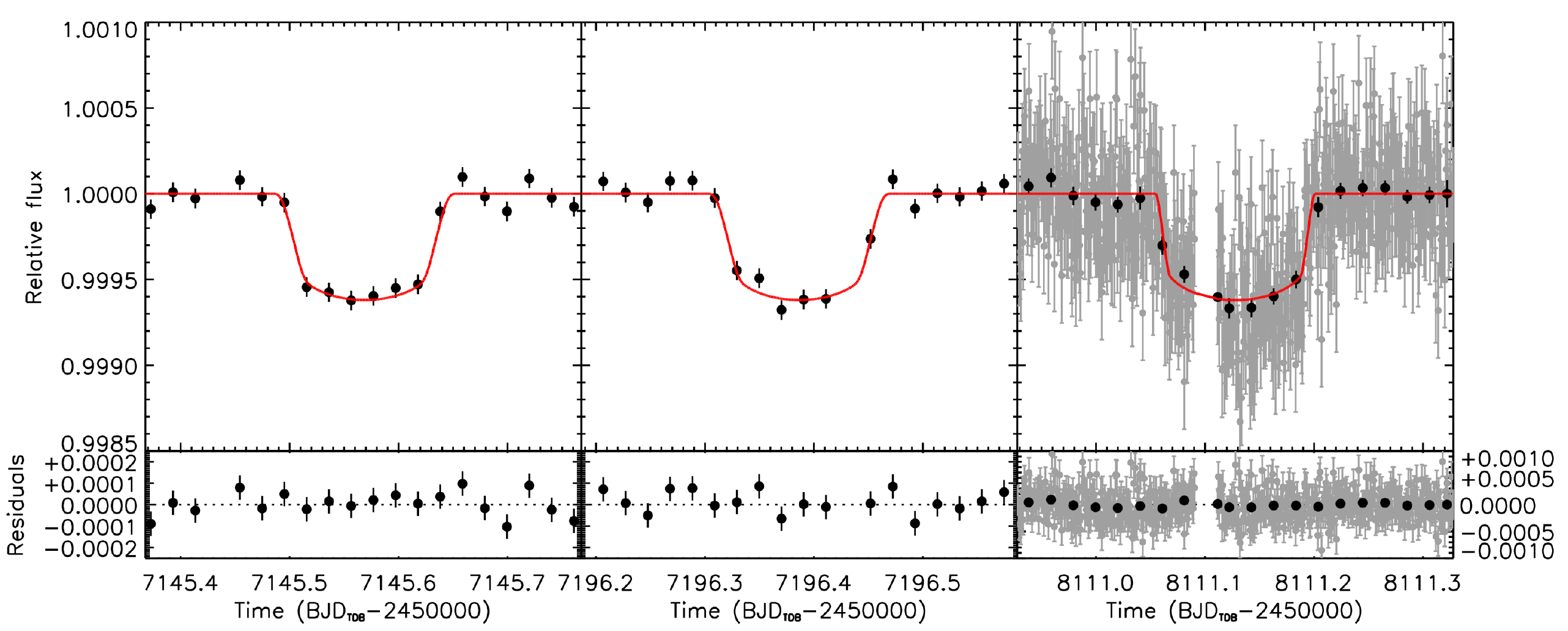}
\caption[]{Normalised flux versus time showing the three transit events. The first two transits were observed in long cadence (29.4 min) and the third one in short cadence (1 min). The short cadence data is shown in grey with binned point overlaid in black. The red solid line indicates our best solution from the combined fit described in Section \ref{combo}. }
\label{fig_transitfit}
\end{center}
\end{figure*}

\begin{figure}
\begin{center}
\includegraphics[width=\linewidth]{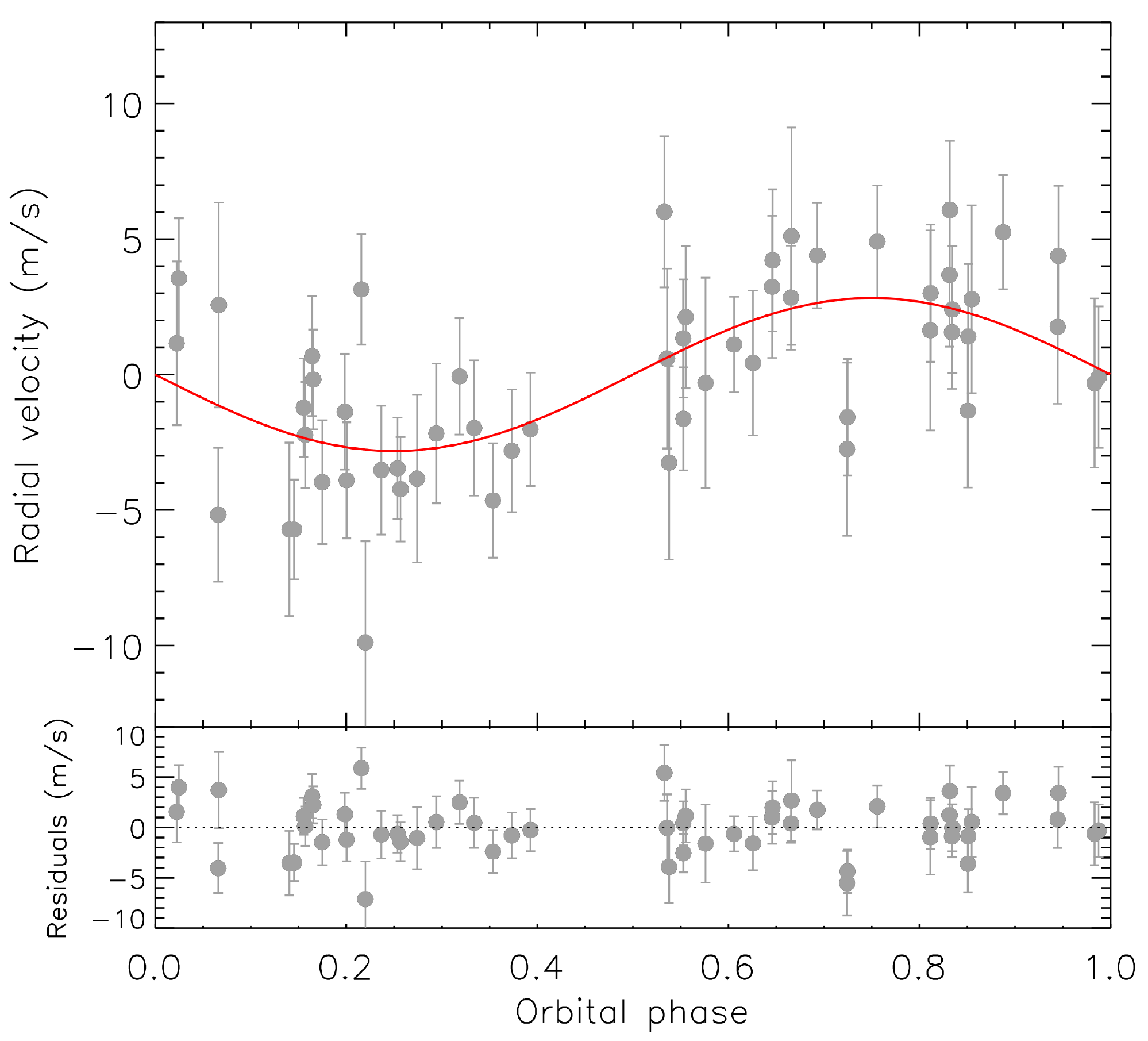}
\includegraphics[width=\linewidth]{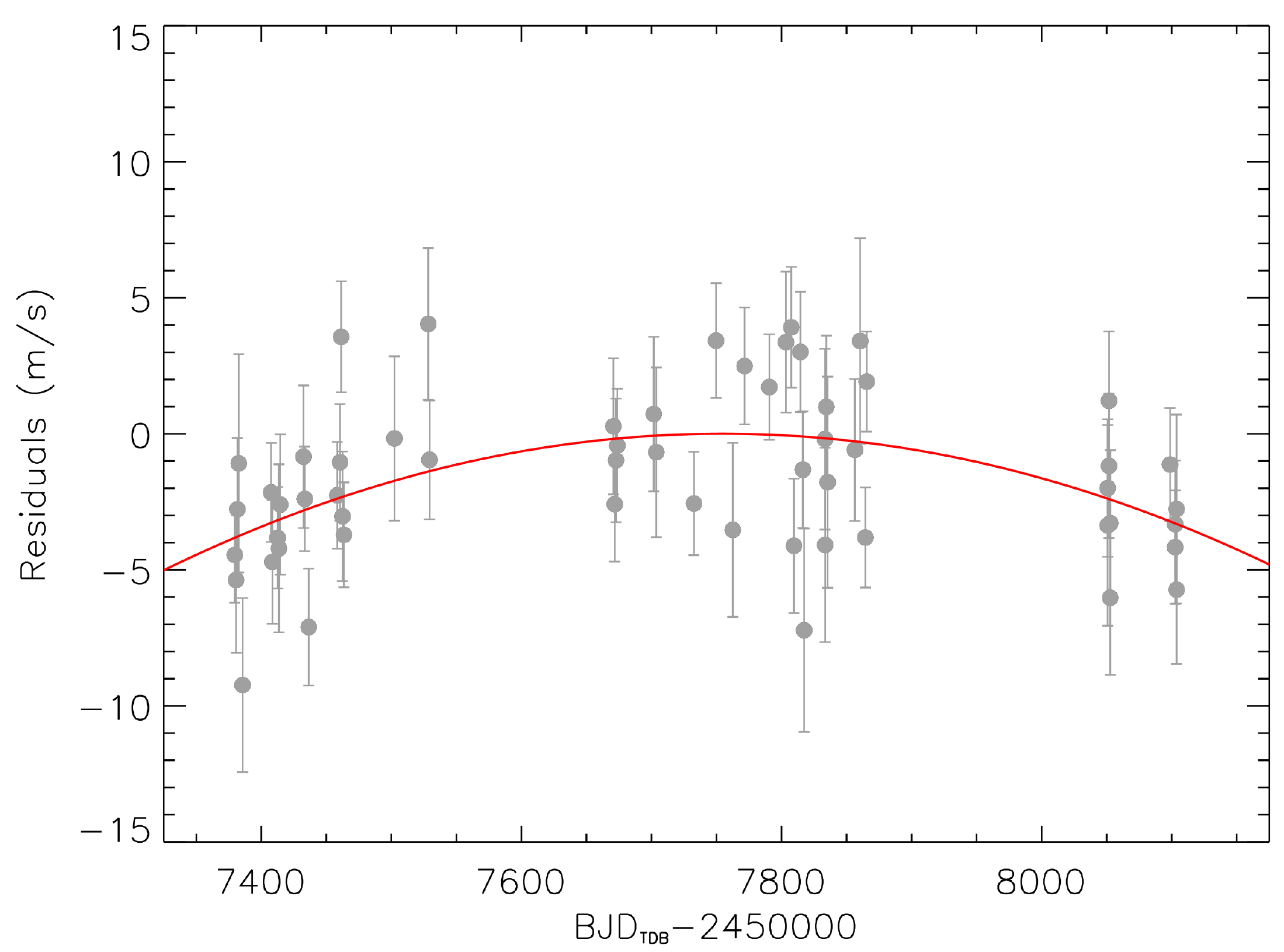}
\caption[]{Top: Radial velocities versus orbital phase after removing the quadratic trend. Transits occur at phase 0/1. The red line indicates the best orbital solution as a result of the combined fit. The bottom panel represent the residuals after removing both the trend and the Keplerian solution. Bottom: Radial velocities versus time after removing the best-fit Keplerian model for \thisplanet. The red line indicates the quadratic trend.}
\label{fig_rvfit}
\end{center}
\end{figure}

\begin{table}
\centering
\caption{\thisstar\ system parameters from combined fit.}            
\label{tab_combofit}
\begin{tabular}{l l }        
\hline\hline                 
\emph{Stellar parameters}  &  \\
\hline
Kepler limb-darkening coefficient $q_{1}$  &  $0.35_{-0.15}^{+0.19}$ \\
Kepler limb-darkening coefficient $q_{2}$  &  $0.51 \pm 0.34$  \\
Kepler limb-darkening coefficient $u_{1}$  &  $0.57 \pm 0.39$ \\
Kepler limb-darkening coefficient $u_{2}$  &  $-0.01_{-0.36}^{+0.41}$ \\
Systemic velocity  $\gamma$ [\kms] & $29.99984 \pm 0.00059$ \\
Linear term  $\dot\gamma~[\rm m s^{-1} d^{-1}]^a $ & $ 7\mbox{\sc{e}-04} \pm 1.9\mbox{\sc{e}-03} $ \\
Quadratic term  $\ddot\gamma~[\rm m s^{-1} d^{-2}]^a $ & $ -5.40\mbox{\sc{e}-05} \pm 1.68\mbox{\sc{e}-05} $ \\
RV jitter $s_{\rm j}$ & $1.11_{-0.64}^{+0.58}~(<1.39)$ \\ 

%\hline
& \\
\hline
\emph{Transit and orbital parameters}  &  \\
\hline
Orbital period $P$ [d] & $50.818947 \pm 0.000094$ \\
Transit epoch $T_{ \rm c} [\rm BJD_{TDB}-2450000$] & $8111.1274 \pm 0.0012$  \\
Transit duration $T_{\rm 14}$ [d] & $0.1453 \pm 0.0038$  \\
Radius ratio $R_{\rm p}/R_{*}$ & $0.0260_{-0.0010}^{+0.0013}$   \\
Inclination $i$ [deg] & $89.24_{-0.07}^{+0.05}$  \\
$a/R_{*}$ & $64.7_{-2.5}^{+2.4}$  \\
Impact parameter $b$ & $0.84_{-0.06}^{+0.03}$  \\
$\sqrt{e}~\cos{\omega}$ &  $0.03_{-0.23}^{+0.21} $ \\
$\sqrt{e}~\sin{\omega}$  &  $0.08 \pm 0.28 $ \\
Orbital eccentricity $e$  &  $< 0.14$   \\
%Argument of periastron $\omega$ [deg] & $145_{-81}^{+93}$ \\
Radial-velocity semi-amplitude $K$ [\ms] & $2.82 \pm 0.58$ \\
%\hline
& \\
\hline
\multicolumn{2}{l}{\emph{Planetary parameters}} \\
\hline
Planet mass $M_{\rm p} ~[\rm M_\oplus]$  &  $14.8 \pm 3.1$  \\
Planet radius $R_{\rm p} ~[\rm R_\oplus]$  &  $2.41 \pm 0.12$  \\
Planet density $\rho_{\rm p}$ [$\rm g\;cm^{-3}$] &  $5.7_{-1.4}^{+1.6}$  \\
Planet surface gravity log\,$g_{\rm p }$ [cgs] &  $3.4_{-0.11}^{+0.10}$  \\
Orbital semi-major axis $a$ [AU] & $0.2573 \pm 0.0029$   \\
Equilibrium temperature $T_{\rm eq}$ [K]~$^b$  & $470 \pm 10$ \\
\hline       
\hline
\vspace{-0.3cm}
%\footnotetext[2]{\scriptsize black-body equilibrium temperature assuming a null Bond albedo and uniform heat redistribution to the night-side} \\
\end{tabular}
\newline {\footnotesize a: reference time is the average of the RV epochs. b: black-body equilibrium temperature assuming a null Bond albedo and uniform heat redistribution to the night-side}
\end{table}

We simultaneously modeled the {\it K2} photometry and the HARPS-N RVs following the same procedure as described in \citet{Bono14,Bono15}. In short, we used a differential evolution Markov chain Monte Carlo (DE-MCMC) Bayesian method \citep{TerB06,East13}. The transit model of \citet{Mandel02} was computed at the same short-cadence sampling (1 min) as the \emph{K2} measurements during Campaign 16. Since data of Campaign 5 were gathered only in long-cadence mode (29.4 min), we oversampled the transit model at 1 min and then averaged it to the long-cadence samples to compute the likelihood function; this allowed us to overcome the well-known smearing effect due to long integration times on the determination of transit parameters \citep{Kip10}.

We accounted for a light travel time of $\sim 2$~min between the \emph{K2} transit observations which are referred to the planet reference frame and the RVs in the stellar frame, given the relatively large semi-major axis of \thisplanet\ ($\sim 0.25$~AU). The free parameters of our global model are the mid-transit time $T_{\rm c}$, the orbital period $P$, the systemic radial velocity $\gamma$, the RV semi-amplitude $K$, two combinations of eccentricity $e$ and argument of periastron $\omega$ (i.e. $\sqrt{e}\cos\omega$ and $\sqrt{e}\cos\omega$), the RV uncorrelated jitter term $s_{\rm j}$ (e.g., \citealt{Greg05}), the transit duration $T_{14}$, the scaled planetary radius $R_p/R_{\ast}$, the orbital inclination $i$, and the two limb-darkening coefficients $q_1$ and $q_2$, which are related to the coefficients $u_1$ and $u_2$ of the quadratic limb-darkening law \citep{Cla04,Kip13}.
After running a first combined analysis, we noticed a curvature in the residuals of the HARPS-N RVs (see bottom plot Fig.~\ref{fig_rvfit}). We thus decided to include an RV linear ($\dot\gamma$) and quadratic ($\ddot\gamma$) term as free parameters, following the formalism by \citet{Kip11}. The reference time for the quadratic trend was chosen to be the average of the epochs of the RV measurements. We imposed a Gaussian prior on the stellar density as derived in Section \ref{star3} and used uninformative priors on all the RV model parameters. Bounds of [0, 1] and [0, 1[ were adopted for the eccentricity and the limb-darkening parameters \citep{Kip13}, respectively. 

We ran twenty-eight chains, which is twice the number of free parameters of our model. The step directions and sizes for each chain were automatically determined from the other chains following \citet{TerB06}. After discarding the burn-in steps and achieving convergence according to the prescriptions given in \citep{East13}, the medians of the posterior distributions were evaluated as the final parameters, and their 34.13\% quantiles were adopted as the associated $1\sigma$ uncertainties. Fitted and derived parameters are listed in Table~\ref{tab_combofit}. The best-fit models of the transits and RVs are displayed in Figs.~\ref{fig_transitfit} and \ref{fig_rvfit}. 

By combining the derived radius ratio and the RV semi-amplitude with the stellar parameters obtained in Section \ref{star3}, we find that \thisplanet\ has a radius of $R_{\rm p}=2.41 \pm 0.12$\,\rearth, a mass of $M_{\rm p} = 14.8 \pm 3.1$\,\mearth, and thus a density of $5.7_{-1.4}^{+1.6}$\,g\,cm$^{-3}$. The eccentricity is consistent with zero at the current precision.  

If the RV curvature is not due to either a long-term activity variation or instrumental systematics, but instead to the presence of a long-period companion, from the $\dot\gamma$ and $\ddot\gamma$ coefficients we estimate the companion orbital period, RV semi-amplitude, and mass to be $> 4.5$~yr, $>3.8~\rm m s^{-1}$, and $>60$\,\mearth, respectively \citep[e.g. ][]{Kip11}. Without including this curvature, the semi-amplitude and thus planetary mass are slightly lower. However, the values of both analyses are within one sigma and fully consistent with each other.

\section{RV analysis with GP}\label{rv}

%------------------------------------------
\begin{figure*}
\begin{center}
\includegraphics[width=\textwidth]{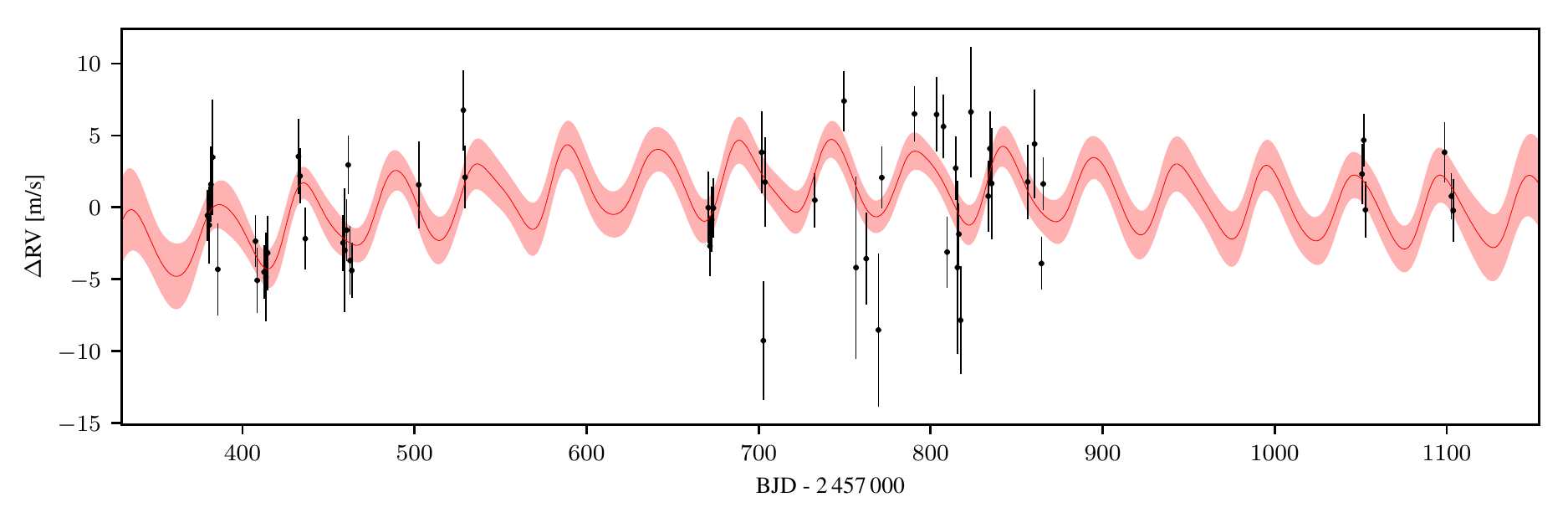}
\caption[]{RVs versus time. The red solid line indicates the GP plus planet model posterior mean; and the shaded region denotes the $\pm\sigma$ posterior uncertainty. Note that the RVs were fitted jointly with activity indicator time series; however, as the GP amplitudes for the latter time series were consistent with zero, the fits for these time series are not plotted here.}
\label{fig:GP_planet_fit}
\end{center}
\end{figure*}
%------------------------------------------

As an independent check on the mass measurement and to compare models, we performed a combined analysis of HARPS-N RVs and spectroscopic activity indices using the Gaussian process (GP) framework introduced in \citet[][hereafter R15]{Raj15} and \citet{Raj16}. This framework was designed specifically to model RVs jointly with activity diagnostics even when simultaneous photometry is not available. It models both activity indices and activity-induced RV variations as a physically-motivated manifestation of a single underlying GP and its derivative. It is able to disentangle stellar signals from planetary ones even in cases where their periods are very close \citep[see e.g.][]{ME16}, whilst at the same time not wrongly identify a planetary signal as stellar activity.

We used R15's framework to derive a joint constraint on the activity component of the RVs and on the mass of planet b. For this analysis, we modelled the \SMW, BIS, FWHM, and RV measurements simultaneously. A GP with a quasi-periodic covariance kernel was used to model stellar activity. For the GP mean function, we considered three models: zero, one, or two non-interacting Keplerian signals in the RVs only. We fixed the first Keplerian signal's period to $50.818947$~d and mid-transit time to $2458111.1274$~d, as informed by the fit of the K2 light curve (see Section \ref{combo}) so that this signal, if detected, would correspond to \thisplanet; we constrained the period of the second Keplerian component (to account for a possible non-transiting planet detectable in the RVs) to lie between $0.1$~d and $1000$~d. For the prior on the orbital eccentricities we used a Beta distribution with parameters $a=0.867$ and $b=3.03$ \citep[see][]{Kip13}, and placed non-informative priors on the remaining orbital elements (uniform) and RV semi-amplitudes (modified Jeffreys). We also placed non-informative priors on all parameters related to the activity components of the GP framework (uniform priors for parameters with known scales and Jeffreys priors for the remaining parameters - for more details see see R15). All parameter and model inference was performed using the \textsc{MultiNest} nested-sampling algorithm \citep{Feroz08,Feroz09,Feroz13}, with $2000$~live points and a sampling efficiency of $0.3$.

We thus computed log model likelihoods (evidences) of $\ln \mathcal{Z}_0=-141.5\pm0.1$, $\ln\mathcal{Z}_1=-135.6\pm0.1$ and $\ln\mathcal{Z}_2=-135.8\pm0.1$ for the 0-, 1- and 2-planet models, respectively. On this basis we concluded that the model corresponding to an RV detection of planet b was favoured decisively over a zero-planet model, with a Bayes factor of $\mathcal{Z}_1/\mathcal{Z}_0>300$. The more complex 2-planet model was not supported with a Bayes factor $\mathcal{Z}_2/\mathcal{Z}_1\sim1.2$.

Using the 1-planet model, we obtained an RV semi-amplitude of $K_b = 2.52\pm0.55$~\ms, and an eccentricity of $e_b=0.08^{+0.11}_{-0.06}$ translating into a planetary mass of $13\pm3$\,\mearth. This value is consistent with the one derived in Section \ref{combo}. The parameters associated to this model can be found in Table \ref{tab_GPfit} and the best fit is plotted in Figure \ref{fig:GP_planet_fit}.

Under the 2-planet model, the posterior distributions for $K_b$ and $e_b$ were consistent with (and essentially identical to) those obtained under the 1-planet model. The periods of the second 'planet' corresponding to nontrivial RV semi-amplitudes were $240^{+40}_{-20}$~d and $880\pm160$~d, where the first one corresponds to a peak in the RV BGLS periodogram and which may be an effect of the seasonal sampling of the data.

Under all three models, we always obtained very broad posterior distributions for the main GP hyper-parameters, indicating that the characteristics of any activity signal present were poorly constrained. In particular, under the favoured 1-planet model, we obtained $P_\textrm{GP}=64^{+57}_{-36}$~d (overall period for the activity signal), $\lambda_\textrm{p}\sim 5.4\pm2.7$ (inverse harmonic complexity, with this inferred value pointing to low harmonic complexity, i.e.\ nearly sinusoidal variability), and $\lambda_\textrm{e}=196^{+72}_{-78}$~d (activity signal evolution time scale). The GP amplitude parameters for the BIS, FWHM, and \SMW\ time series were all smaller than about $10\%$ of the rms variation observed in each series, and indeed smaller than the estimated noise variance for each series. Thus, the GP fit suggests that there is something present in the data that is probably not simply white noise, but also cannot be interpreted as clear evidence of another planet (since the evidence for a 2-planet model is very weak) nor as activity (since no coherent signals show up in the activity indicators). Instead, these RV signals accounted for by the GP may be due to one or multiple undetected planets, instrumental or observational effects, etc.

For completeness, we have also fitted the RVs with a Keplerian without the use of a GP. Our conclusions about the planet parameters were virtually identical to and entirely consistent with those from the 1-planet plus GP model. Additionally, we ran the same analysis using a uniform prior for the eccentricity rather than the prior suggested by \citet{Kip13}. Again, the results were entirely consistent and thus insensitive to the choice of eccentricity prior.

%------------------------------------------
% Table: fit results
%------------------------------------------
\begin{table}
  \centering
  \caption[Posterior summaries: GP plus 1-planet model]{Posterior probability summaries for the main GP covariance and mean function parameters of interest (one planet plus activity model) for our favoured fit to the \thisstar\ data. The Keplerian orbital parameters are as defined in \citet{Sea11}, while the GP hyper-parameters are as defined in R15.}

    \begin{tabular}{l l}
\hline\hline                 
\emph{GP parameters}  &  \\
\hline
GP RV semi-amplitude $K_\textrm{GP}$ [\ms] & $2.68\pm0.52$ \\
GP period $P$ [d] &	$64_{-37}^{+57}$ \\
GP inv.\ harmonic complexity $\lambda_\textrm{p}$ & $5.4_{-2.6}^{+2.9}$ \\
GP evolution time-scale $\lambda_\textrm{e}$ [d] & $196^{+72}_{-78}$ \\
& \\
\hline
\emph{Planet parameters}  &  \\
\hline
System velocity $\gamma$ [\kms] & $-29.837\pm0.0015$ \\
RV semi-amplitude $K_b$ [\ms] & $2.52_{-0.52}^{+0.57}$ \\
Period $P_b$ [d] & $50.818947$ (fixed) \\
Eccentricity $e_b$ & $0.08^{+0.11}_{-0.06}$ \\
Periapsis longitude $\omega_b$ & $0.97\pi^{+0.61\pi}_{-0.58\pi}$ \\
Transit epoch $\textrm{T}_\textrm{c}$ (BJD)	& $2458111.1274$ (fixed) \\
Mass $M_b$ [\mearth] & $13\pm3$ \\
Mean density $\rho_b$ [g~cm$^{-3}$] & $5.1\pm1.2$ \\
\hline\hline
% Then the results using the uniform prior on eccentricity. Basically everything comes out identical to the previous case (e.g. RV semi-amplitude 2.52 -0.57/+0.59 m/s); the eccentricity does seem to shift a tiny bit upwards to 0.13 (-0.09/+0.15) but this is essentially identical to the previous fit.

%    \begin{tabular}{cccc}
%\hline
%Parameter  &	Description							& Median 					& $\pm\sigma$ \\
%\hline
%$K_\textrm{GP}$ 		&	GP RV semi-amplitude						& $2.68$~\ms 					& $\pm0.52$~\ms \\
%$P$ 		&				GP period			& $64$~d 					& $_{-37}^{+57}$~d \\
%$\lambda_\textrm{p}$ &		GP inv.\ harmonic complexity					& $5.4$						& $_{-2.6}^{+2.9}$ \\
%$\lambda_\textrm{e}$  &		GP evolution time-scale				& $196$~d 						& $^{+72}_{-78}$~d \\
%\hline
%$\gamma$ &	System velocity 			&$-29.837$ \kms				& $\pm1.5$ \ms \\
%$K_b$	&		RV semi-amplitude 			& $2.58$ \ms					& $_{-0.53}^{+0.56}$~\ms \\
%$P_b$	&		Period				& $50.816724\textrm{ d }$		& (fixed) \\
%$e_b$ 	&		Eccentricity			& $0.08$						& $^{+0.11}_{-0.06}$ \\
%$\omega_b$ 	& Periapsis longitude		& $0.97\pi$						& $^{+0.61}_{-0.58}\pi$ \\
%$\textrm{T}_\textrm{tr,b}$ 	& Transit mid-point time (BJD)	& $2457145.56666$~d		& (fixed) \\
%$M_b$ & Mass & $13.6$~\mearth & $\pm2.9$~\mearth \\
%$\rho_b$  & Mean density & $5.5$~g~cm$^{-3}$ & $\pm1.5$~g~cm$^{-3}$ \\
%\hline

    \end{tabular}%
  \label{tab_GPfit}%
\end{table}%
%------------------------------------------

\section{Discussion and Conclusions}\label{concl}

\begin{figure*}
\begin{center}
\includegraphics[width=\textwidth]{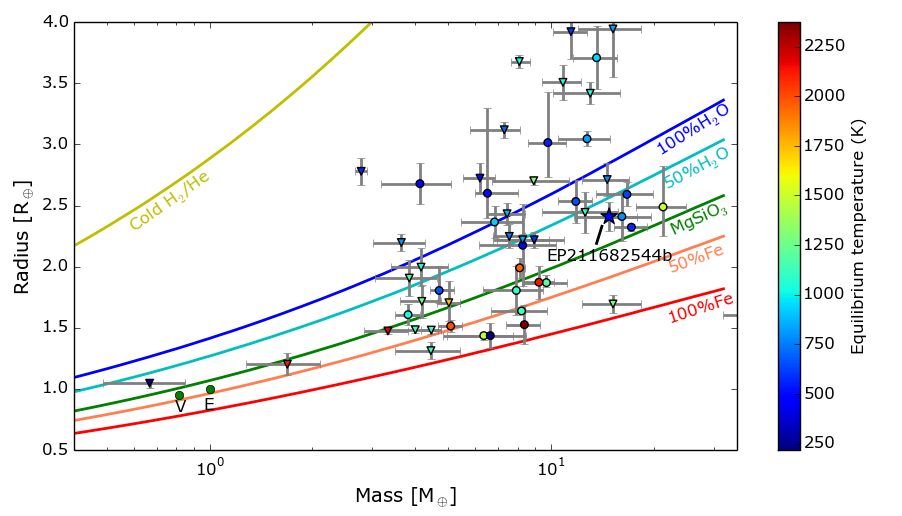}
\caption[]{Mass-radius diagram of all planets smaller than $4$\,\rearth with a mass precision better than 30\% (using exoplanet.eu data). The points are colour-coded according to their equilibrium temperature (assuming $f=1$ and albedo $A=0$). The green dots bottom left represent Venus and Earth. The solid lines show planetary interior models for different compositions, top to bottom: Cold H$_2$/He, 100\% H$_2$O, 50\% H$_2$O, 100\% MgSiO$_3$, 50\% Fe, 100\% Fe. The large star represents \thisplanet, dots are planets where the mass was obtained via RV, and triangles are planets where the mass was via TTV.}
\label{fig_mr}
\end{center}
\end{figure*}

We used high-resolution spectroscopy to characterise \thisstar\ and determine the mass of its orbiting planet, \thisplanet. A combined analysis of the precise RVs and the {\it K2} lightcurve reveals that this planet has an orbital period of $50.818947\pm 0.000094$\,days, a radius of $2.41\pm0.12$\,\rearth, and a mass of $14.8\pm3.1$\,\mearth. 

Stellar contamination in the RVs can complicate the analysis and influence the planetary mass determination. Despite \thisstar\ being a quiet star, we ran a GP analysis of the RVs and the standard activity indicators. The mass determination agrees with the one from the combined fit. The activity indicators showed no significant variation and the GP hyperparameters were poorly constrained. As shown by the GP analysis, there are time-correlated signals in the RVs that could not be ascribed to planet b and that are not represented in the time series of the standard activity indicators. A two planet model, however, was not favoured for these data.

Figure \ref{fig_mr} shows the mass-radius diagram for all small planets ($R_p<4$\,\rearth) with a planetary mass determined with a precision better than 30\%\footnote{Data from \url{www.exoplanet.eu} \citep{Sch11}}. Overplotted are radius-mass relations representing different planet compositions \citep{Zeng13,Zeng16}. \thisplanet\ has a bulk density in between that of an Earth-like rocky planet ($32.5\%$ Fe/Ni-metal + $67.5\%$ Mg-silicates-rock) and that of a pure-100\% H$_2$O planet. Specifically, the median value of its density estimate ($\rho_p=5.7\pm1.5$\,g\,cm$^{-3}$) implies that it most likely contains an equivalent amount of ices compared to rocks, that is, 50\% ices and 50\% rock+metal. This proportion is expected from the abundance ratio of major planet-building elements, including Fe, Ni, Mg, Si, O, C, N, in a solar-like nebula.

Its mass of $14.8$ Earth masses together with its estimated composition (half rock+metal and half ices) suggest that \thisplanet\ likely formed in a similar way as the cores of giant planets in our own Solar System (Jupiter, Saturn, Uranus, Neptune), but for some reason, it did not accrete much gas. This would require its initial formation beyond or near the snowline in its own system, followed by subsequent inward migration to its current position of $\sim0.25$\,AU from its host star. Considering the smaller mass ($0.88$\,\msun) and luminosity ($0.55$\,\lsun) of its host star compared to the Sun, the snowline position in this particular system should be somewhat closer than it is for the Solar System. The position of the snowline in our own Solar System lies around 3\,AU, right in the middle of the asteroid belt \citep[e.g.]{Haya81,Podo04,Mart12}. Its location can also move inward with time \citep{Sas00,Sat16}. Naively scaling by the luminosity of the central star, the snowline for this system is expected to be around 2\,AU.

\begin{figure}
\begin{center}
\includegraphics[width=\linewidth]{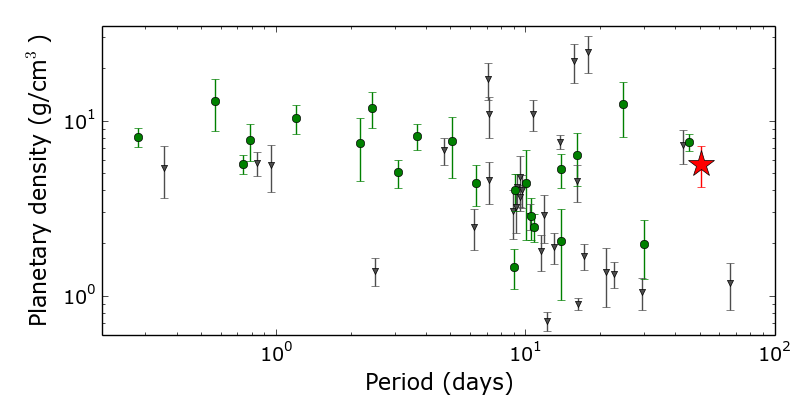}
\caption[]{Planetary bulk density versus orbital period for the same planets as in Figure \ref{fig_mr}. Green dots indicate the planets which mass was determined via radial velocity and grey triangles the TTV determined planets. The red star is \thisplanet.}
\label{fig_densper}
\end{center}
\end{figure}

In terms of its mass and radius, \thisplanet\ is very similar to Kepler-131\,b \citep{Marcy14} and HD106315\,b \citep{Barr17,Cros17}, but the longer period of \thisplanet\ makes it significantly cooler ($T_{eq} = 470$\,K). In fact, \thisplanet\ currently has the longest period of all small planets ($R_p<4$\,\rearth) in Figure \ref{fig_mr} where the mass was determined via RVs (see Figure \ref{fig_densper}). The only longer-period planet in this figure is Kepler-289\,d \citep{Schm14} with a period of $\sim66$\,days. However, its mass was determined via transit time variations.

In the mass-radius diagram, \thisplanet\ lies among a group of exoplanets in between 2 and 3 Earth radii with similar masses (5-20 Earth masses) and similar insolation/surface-equilibrium-temperatures. This entire group of exoplanets correspond to a peak in the planet size distribution \citep{Zeng18} above the recently-discovered exoplanet radius gap around 2 Earth radii in the Kepler planet data \citep[e.g.][]{Zeng17a,Zeng17b,Ful17,Ful18,VanE18,Ber18,Mayo18,Thom18}. It means that these kind of icy cores (which must also contain rock+metal) form quite easily and favorably among solar-like stars. If correct, planets between 2 and 3 Earth radii should be in the same mass range. Simulations predict that TESS will discover 561 planets in this radius range, with about half orbiting stars brighter than $V=12$ \citep{Barc18}. Future RV observations of TESS planets could thus confirm this theory.

\section*{Acknowledgements}

We thank the anonymous referee for a prompt report.

The HARPS-N project has been funded by the Prodex Program of the Swiss Space Office (SSO), the Harvard University Origins of Life Initiative (HUOLI), the Scottish Universities Physics Alliance (SUPA), the University of Geneva, the Smithsonian Astrophysical Observatory (SAO), and the Italian National Astrophysical Institute (INAF), the University of St Andrews, Queen's University Belfast, and the University of Edinburgh.

The research leading to these results received funding from the European Union Seventh Framework Programme (FP7/2007- 2013) under grant agreement number 313014 (ETAEARTH).
VMR acknowledges the Royal Astronomical Society and Emmanuel College, Cambridge, for financial support.
This work was performed in part under contract with the California Institute of Technology (Caltech)/Jet Propulsion Laboratory (JPL) funded by NASA through the Sagan Fellowship Program executed by the NASA Exoplanet Science Institute (AV, RDH). 
LM acknowledges the support by INAF/Frontiera through the "Progetti Premiali" funding scheme of the Italian Ministry of Education, University, and Research.
ACC acknowledges support from STFC consolidated grant number ST/R000824/1.
DWL acknowledges partial support from the Kepler mission under NASA Cooperative Agreement NNX13AB58A with the Smithsonian Astrophysical Observatory. 
CAW acknowledges support by STFC grant ST/P000312/1.
Some of this work has been carried out within the framework of the NCCR PlanetS, supported by the Swiss National Science Foundation.
This publication was made possible through the support of a grant from the John Templeton Foundation. The opinions expressed are those of the authors and do not necessarily reflect the views of the John Templeton Foundation. 

This material is based upon work supported by the National Aeronautics and Space Administration under grants No. NNX15AC90G and NNX17AB59G issued through the Exoplanets Research Program.
This research has made use of the SIMBAD database, operated at CDS, Strasbourg, France, NASA's Astrophysics Data System and the NASA Exoplanet Archive, which is operated by the California Institute of Technology, under contract with the National Aeronautics and Space Administration under the Exoplanet Exploration Program.
Based on observations made with the Italian Telescopio Nazionale Galileo (TNG) operated on the island of La Palma by the Fundacion Galileo Galilei of the INAF (Istituto Nazionale di Astrofisica) at the Spanish Observatorio del Roque de los Muchachos of the Instituto de Astrofisica de Canarias.
This paper includes data collected by the \emph{K2}\ mission. Funding for the \emph{K2}\ mission is provided by the NASA Science Mission directorate. Some of the data presented in this paper were obtained from the Mikulski Archive for Space Telescopes (MAST). STScI is operated by the Association of Universities for Research in Astronomy, Inc., under NASA contract NAS5-26555. Support for MAST for non-HST data is provided by the NASA Office of Space Science via grant NNX13AC07G and by other grants and contracts. \thisstar\ was observed as part of the following Guest Programmes: GO5007\_LC (PI: Winn), GO5029\_LC (PI: Charbonneau), GO5033\_LC (PI: Howard), GO5104\_LC (PI: Dragomir), GO5106\_LC (PI: Jackson), GO5060\_LC (PI: Coughlin), GO16009\_LC (PI: Charbonneau), GO16011\_LC (PI: Fabrycky), GO16015\_LC (PI: Boyajian), GO16020\_LC (PI: Adams), GO16021\_LC (PI: Howard), GO16101\_LC (PI: Winn), GO16009\_SC (PI: Charbonneau), GO16015\_SC (PI: Boyajian), and GO16101\_SC (PI: Winn).

%%%%%%%%%%%%%%%%%%%%%%%%%%%%%%%%%%%%%%%%%%%%%%%%%%



\bibliographystyle{mnras}
\bibliography{References} % if your bibtex file is called example.bib

\begin{thebibliography}{}
\makeatletter
\relax
\def\mn@urlcharsother{\let\do\@makeother \do\$\do\&\do\#\do\^\do\_\do\%\do\~}
\def\mn@doi{\begingroup\mn@urlcharsother \@ifnextchar [ {\mn@doi@}
  {\mn@doi@[]}}
\def\mn@doi@[#1]#2{\def\@tempa{#1}\ifx\@tempa\@empty \href
  {http://dx.doi.org/#2} {doi:#2}\else \href {http://dx.doi.org/#2} {#1}\fi
  \endgroup}
\def\mn@eprint#1#2{\mn@eprint@#1:#2::\@nil}
\def\mn@eprint@arXiv#1{\href {http://arxiv.org/abs/#1} {{\tt arXiv:#1}}}
\def\mn@eprint@dblp#1{\href {http://dblp.uni-trier.de/rec/bibtex/#1.xml}
  {dblp:#1}}
\def\mn@eprint@#1:#2:#3:#4\@nil{\def\@tempa {#1}\def\@tempb {#2}\def\@tempc
  {#3}\ifx \@tempc \@empty \let \@tempc \@tempb \let \@tempb \@tempa \fi \ifx
  \@tempb \@empty \def\@tempb {arXiv}\fi \@ifundefined
  {mn@eprint@\@tempb}{\@tempb:\@tempc}{\expandafter \expandafter \csname
  mn@eprint@\@tempb\endcsname \expandafter{\@tempc}}}

\bibitem[\protect\citeauthoryear{{Baranne} et~al.,}{{Baranne}
  et~al.}{1996}]{Bara96}
{Baranne} A.,  et~al., 1996, \aaps, \href
  {http://adsabs.harvard.edu/abs/1996A%26AS..119..373B} {119, 373}

\bibitem[\protect\citeauthoryear{{Barclay}, {Pepper}  \& {Quintana}}{{Barclay}
  et~al.}{2018}]{Barc18}
{Barclay} T.,  {Pepper} J.,   {Quintana} E.~V.,  2018, preprint, \href
  {http://adsabs.harvard.edu/abs/2018arXiv180405050B} {} (\mn@eprint {arXiv}
  {1804.05050})

\bibitem[\protect\citeauthoryear{{Barros} et~al.,}{{Barros}
  et~al.}{2017}]{Barr17}
{Barros} S.~C.~C.,  et~al., 2017, \mn@doi [\aap] {10.1051/0004-6361/201731276},
  \href {http://adsabs.harvard.edu/abs/2017A%26A...608A..25B} {608, A25}

\bibitem[\protect\citeauthoryear{{Berger}, {Huber}, {Gaidos}  \& {van
  Saders}}{{Berger} et~al.}{2018}]{Ber18}
{Berger} T.~A.,  {Huber} D.,  {Gaidos} E.,   {van Saders} J.~L.,  2018,
  preprint, \href {http://adsabs.harvard.edu/abs/2018arXiv180500231B} {}
  (\mn@eprint {arXiv} {1805.00231})

\bibitem[\protect\citeauthoryear{{Bonomo} et~al.,}{{Bonomo}
  et~al.}{2014}]{Bono14}
{Bonomo} A.~S.,  et~al., 2014, \mn@doi [\aap] {10.1051/0004-6361/201424617},
  \href {http://adsabs.harvard.edu/abs/2014A%26A...572A...2B} {572, A2}

\bibitem[\protect\citeauthoryear{{Bonomo} et~al.,}{{Bonomo}
  et~al.}{2015}]{Bono15}
{Bonomo} A.~S.,  et~al., 2015, \mn@doi [\aap] {10.1051/0004-6361/201323042},
  \href {http://adsabs.harvard.edu/abs/2015A%26A...575A..85B} {575, A85}

\bibitem[\protect\citeauthoryear{{Bressan}, {Marigo}, {Girardi}, {Salasnich},
  {Dal Cero}, {Rubele}  \& {Nanni}}{{Bressan} et~al.}{2012}]{Bre12}
{Bressan} A.,  {Marigo} P.,  {Girardi} L.,  {Salasnich} B.,  {Dal Cero} C.,
  {Rubele} S.,   {Nanni} A.,  2012, \mn@doi [\mnras]
  {10.1111/j.1365-2966.2012.21948.x}, \href
  {http://adsabs.harvard.edu/abs/2012MNRAS.427..127B} {427, 127}

\bibitem[\protect\citeauthoryear{{Buchhave} et~al.,}{{Buchhave}
  et~al.}{2012}]{Buch12}
{Buchhave} L.~A.,  et~al., 2012, \mn@doi [\nat] {10.1038/nature11121}, \href
  {http://adsabs.harvard.edu/abs/2012Natur.486..375B} {486, 375}

\bibitem[\protect\citeauthoryear{{Buchhave} et~al.,}{{Buchhave}
  et~al.}{2014}]{Buch14}
{Buchhave} L.~A.,  et~al., 2014, \mn@doi [\nat] {10.1038/nature13254}, \href
  {http://adsabs.harvard.edu/abs/2014Natur.509..593B} {509, 593}

\bibitem[\protect\citeauthoryear{{Claret}}{{Claret}}{2004}]{Cla04}
{Claret} A.,  2004, \mn@doi [\aap] {10.1051/0004-6361:20041673}, \href
  {http://adsabs.harvard.edu/abs/2004A%26A...428.1001C} {428, 1001}

\bibitem[\protect\citeauthoryear{{Cosentino} et~al.,}{{Cosentino}
  et~al.}{2012}]{Cos12}
{Cosentino} R.,  et~al., 2012, in Ground-based and Airborne Instrumentation for
  Astronomy IV. p. 84461V, \mn@doi{10.1117/12.925738}

\bibitem[\protect\citeauthoryear{{Crossfield} et~al.,}{{Crossfield}
  et~al.}{2017}]{Cros17}
{Crossfield} I.~J.~M.,  et~al., 2017, \mn@doi [\aj] {10.3847/1538-3881/aa6e01},
  \href {http://adsabs.harvard.edu/abs/2017AJ....153..255C} {153, 255}

\bibitem[\protect\citeauthoryear{{Dressing} et~al.,}{{Dressing}
  et~al.}{2017}]{Dre17}
{Dressing} C.~D.,  et~al., 2017, \mn@doi [\aj] {10.3847/1538-3881/aa89f2},
  \href {http://adsabs.harvard.edu/abs/2017AJ....154..207D} {154, 207}

\bibitem[\protect\citeauthoryear{{Eastman}, {Gaudi}  \& {Agol}}{{Eastman}
  et~al.}{2013}]{East13}
{Eastman} J.,  {Gaudi} B.~S.,   {Agol} E.,  2013, \mn@doi [\pasp]
  {10.1086/669497}, \href {http://adsabs.harvard.edu/abs/2013PASP..125...83E}
  {125, 83}

\bibitem[\protect\citeauthoryear{{Feroz} \& {Hobson}}{{Feroz} \&
  {Hobson}}{2008}]{Feroz08}
{Feroz} F.,  {Hobson} M.~P.,  2008, \mn@doi [\mnras]
  {10.1111/j.1365-2966.2007.12353.x}, \href
  {http://adsabs.harvard.edu/abs/2008MNRAS.384..449F} {384, 449}

\bibitem[\protect\citeauthoryear{{Feroz}, {Hobson}  \& {Bridges}}{{Feroz}
  et~al.}{2009}]{Feroz09}
{Feroz} F.,  {Hobson} M.~P.,   {Bridges} M.,  2009, \mn@doi [\mnras]
  {10.1111/j.1365-2966.2009.14548.x}, \href
  {http://adsabs.harvard.edu/abs/2009MNRAS.398.1601F} {398, 1601}

\bibitem[\protect\citeauthoryear{{Feroz}, {Hobson}, {Cameron}  \&
  {Pettitt}}{{Feroz} et~al.}{2013}]{Feroz13}
{Feroz} F.,  {Hobson} M.~P.,  {Cameron} E.,   {Pettitt} A.~N.,  2013, ArXiv
  e-prints: 1306.2144, \href
  {http://adsabs.harvard.edu/abs/2013arXiv1306.2144F} {}

\bibitem[\protect\citeauthoryear{{Flower}}{{Flower}}{1996}]{Flo96}
{Flower} P.~J.,  1996, \mn@doi [\apj] {10.1086/177785}, \href
  {http://adsabs.harvard.edu/abs/1996ApJ...469..355F} {469, 355}

\bibitem[\protect\citeauthoryear{{Fulton} \& {Petigura}}{{Fulton} \&
  {Petigura}}{2018}]{Ful18}
{Fulton} B.~J.,  {Petigura} E.~A.,  2018, preprint, \href
  {http://adsabs.harvard.edu/abs/2018arXiv180501453F} {} (\mn@eprint {arXiv}
  {1805.01453})

\bibitem[\protect\citeauthoryear{{Fulton} et~al.,}{{Fulton}
  et~al.}{2017}]{Ful17}
{Fulton} B.~J.,  et~al., 2017, \mn@doi [\aj] {10.3847/1538-3881/aa80eb}, \href
  {http://adsabs.harvard.edu/abs/2017AJ....154..109F} {154, 109}

\bibitem[\protect\citeauthoryear{{Gaia Collaboration} et~al.,}{{Gaia
  Collaboration} et~al.}{2016}]{Gaia16}
{Gaia Collaboration} et~al., 2016, \mn@doi [\aap]
  {10.1051/0004-6361/201629272}, \href
  {http://adsabs.harvard.edu/abs/2016A%26A...595A...1G} {595, A1}

\bibitem[\protect\citeauthoryear{{Gaia Collaboration}, {Brown}, {Vallenari},
  {Prusti}, {de Bruijne}, {Babusiaux}  \& {Bailer-Jones}}{{Gaia Collaboration}
  et~al.}{2018}]{Gaia2}
{Gaia Collaboration} {Brown} A.~G.~A.,  {Vallenari} A.,  {Prusti} T.,  {de
  Bruijne} J.~H.~J.,  {Babusiaux} C.,   {Bailer-Jones} C.~A.~L.,  2018,
  preprint, \href {http://adsabs.harvard.edu/abs/2018arXiv180409365G} {}
  (\mn@eprint {arXiv} {1804.09365})

\bibitem[\protect\citeauthoryear{{Green} et~al.,}{{Green}
  et~al.}{2018}]{Green18}
{Green} G.~M.,  et~al., 2018, \mn@doi [\mnras] {10.1093/mnras/sty1008}, \href
  {http://adsabs.harvard.edu/abs/2018MNRAS.478..651G} {478, 651}

\bibitem[\protect\citeauthoryear{{Gregory}}{{Gregory}}{2005}]{Greg05}
{Gregory} P.~C.,  2005, \mn@doi [\apj] {10.1086/432594}, \href
  {http://adsabs.harvard.edu/abs/2005ApJ...631.1198G} {631, 1198}

\bibitem[\protect\citeauthoryear{{Hayashi}}{{Hayashi}}{1981}]{Haya81}
{Hayashi} C.,  1981, \mn@doi [Progress of Theoretical Physics Supplement]
  {10.1143/PTPS.70.35}, \href
  {http://adsabs.harvard.edu/abs/1981PThPS..70...35H} {70, 35}

\bibitem[\protect\citeauthoryear{{Kipping}}{{Kipping}}{2010}]{Kip10}
{Kipping} D.~M.,  2010, \mn@doi [\mnras] {10.1111/j.1365-2966.2010.17242.x},
  \href {http://adsabs.harvard.edu/abs/2010MNRAS.408.1758K} {408, 1758}

\bibitem[\protect\citeauthoryear{{Kipping}}{{Kipping}}{2013}]{Kip13}
{Kipping} D.~M.,  2013, \mn@doi [\mnras] {10.1093/mnrasl/slt075}, \href
  {http://adsabs.harvard.edu/abs/2013MNRAS.434L..51K} {434, L51}

\bibitem[\protect\citeauthoryear{{Kipping} et~al.,}{{Kipping}
  et~al.}{2011}]{Kip11}
{Kipping} D.~M.,  et~al., 2011, \mn@doi [\aj] {10.1088/0004-6256/142/3/95},
  \href {http://adsabs.harvard.edu/abs/2011AJ....142...95K} {142, 95}

\bibitem[\protect\citeauthoryear{{Kurucz}}{{Kurucz}}{1993}]{Kur93}
{Kurucz} R.,  1993, ATLAS9 Stellar Atmosphere Programs and 2 km/s grid.~Kurucz
  CD-ROM No.~13.~ Cambridge, Mass.: Smithsonian Astrophysical Observatory,
  1993., \href {http://adsabs.harvard.edu/abs/1993KurCD..13.....K} {13}

\bibitem[\protect\citeauthoryear{{Malavolta} et~al.,}{{Malavolta}
  et~al.}{2017}]{Mal17}
{Malavolta} L.,  et~al., 2017, \mn@doi [\aj] {10.3847/1538-3881/aa6897}, \href
  {http://adsabs.harvard.edu/abs/2017AJ....153..224M} {153, 224}

\bibitem[\protect\citeauthoryear{{Mamajek} \& {Hillenbrand}}{{Mamajek} \&
  {Hillenbrand}}{2008}]{Mama08}
{Mamajek} E.~E.,  {Hillenbrand} L.~A.,  2008, \mn@doi [\apj] {10.1086/591785},
  \href {http://adsabs.harvard.edu/abs/2008ApJ...687.1264M} {687, 1264}

\bibitem[\protect\citeauthoryear{{Mandel} \& {Agol}}{{Mandel} \&
  {Agol}}{2002}]{Mandel02}
{Mandel} K.,  {Agol} E.,  2002, \mn@doi [\apjl] {10.1086/345520}, \href
  {http://adsabs.harvard.edu/abs/2002ApJ...580L.171M} {580, L171}

\bibitem[\protect\citeauthoryear{{Marcy} et~al.,}{{Marcy}
  et~al.}{2014}]{Marcy14}
{Marcy} G.~W.,  et~al., 2014, \mn@doi [\apjs] {10.1088/0067-0049/210/2/20},
  \href {http://adsabs.harvard.edu/abs/2014ApJS..210...20M} {210, 20}

\bibitem[\protect\citeauthoryear{{Martin} \& {Livio}}{{Martin} \&
  {Livio}}{2012}]{Mart12}
{Martin} R.~G.,  {Livio} M.,  2012, \mn@doi [\mnras]
  {10.1111/j.1745-3933.2012.01290.x}, \href
  {http://adsabs.harvard.edu/abs/2012MNRAS.425L...6M} {425, L6}

\bibitem[\protect\citeauthoryear{Mayo et~al.,}{Mayo et~al.}{2018a}]{Mayo18b}
Mayo A.~W.,  et~al., 2018a, {275 Candidates and 149 Validated Planets Orbiting
  Bright Stars in K2 Campaigns 0-10}, \mn@doi{10.5281/zenodo.1164791}, \url
  {https://doi.org/10.5281/zenodo.1164791}

\bibitem[\protect\citeauthoryear{{Mayo} et~al.,}{{Mayo} et~al.}{2018b}]{Mayo18}
{Mayo} A.~W.,  et~al., 2018b, \mn@doi [\aj] {10.3847/1538-3881/aaadff}, \href
  {http://adsabs.harvard.edu/abs/2018AJ....155..136M} {155, 136}

\bibitem[\protect\citeauthoryear{{Mortier}, {Sousa}, {Adibekyan}, {Brand{\~a}o}
   \& {Santos}}{{Mortier} et~al.}{2014}]{ME14}
{Mortier} A.,  {Sousa} S.~G.,  {Adibekyan} V.~Z.,  {Brand{\~a}o} I.~M.,
  {Santos} N.~C.,  2014, \mn@doi [\aap] {10.1051/0004-6361/201424537}, \href
  {http://adsabs.harvard.edu/abs/2014A%26A...572A..95M} {572, A95}

\bibitem[\protect\citeauthoryear{{Mortier}, {Faria}, {Correia}, {Santerne}  \&
  {Santos}}{{Mortier} et~al.}{2015}]{ME15}
{Mortier} A.,  {Faria} J.~P.,  {Correia} C.~M.,  {Santerne} A.,   {Santos}
  N.~C.,  2015, \mn@doi [\aap] {10.1051/0004-6361/201424908}, \href
  {http://adsabs.harvard.edu/abs/2015A%26A...573A.101M} {573, A101}

\bibitem[\protect\citeauthoryear{{Mortier} et~al.,}{{Mortier}
  et~al.}{2016}]{ME16}
{Mortier} A.,  et~al., 2016, \mn@doi [\aap] {10.1051/0004-6361/201526905},
  \href {http://adsabs.harvard.edu/abs/2016A%26A...585A.135M} {585, A135}

\bibitem[\protect\citeauthoryear{{Morton}}{{Morton}}{2012}]{Morton12}
{Morton} T.~D.,  2012, \mn@doi [\apj] {10.1088/0004-637X/761/1/6}, \href
  {http://adsabs.harvard.edu/abs/2012ApJ...761....6M} {761, 6}

\bibitem[\protect\citeauthoryear{{Morton}, {Bryson}, {Coughlin}, {Rowe},
  {Ravichandran}, {Petigura}, {Haas}  \& {Batalha}}{{Morton}
  et~al.}{2016}]{Morton16}
{Morton} T.~D.,  {Bryson} S.~T.,  {Coughlin} J.~L.,  {Rowe} J.~F.,
  {Ravichandran} G.,  {Petigura} E.~A.,  {Haas} M.~R.,   {Batalha} N.~M.,
  2016, \mn@doi [\apj] {10.3847/0004-637X/822/2/86}, \href
  {http://adsabs.harvard.edu/abs/2016ApJ...822...86M} {822, 86}

\bibitem[\protect\citeauthoryear{{Moya}, {Zuccarino}, {Chaplin}  \&
  {Davies}}{{Moya} et~al.}{2018}]{Moya18}
{Moya} A.,  {Zuccarino} F.,  {Chaplin} W.~J.,   {Davies} G.~R.,  2018,
  preprint, \href {http://adsabs.harvard.edu/abs/2018arXiv180606574M} {}
  (\mn@eprint {arXiv} {1806.06574})

\bibitem[\protect\citeauthoryear{{Noyes}, {Hartmann}, {Baliunas}, {Duncan}  \&
  {Vaughan}}{{Noyes} et~al.}{1984}]{Noy84b}
{Noyes} R.~W.,  {Hartmann} L.~W.,  {Baliunas} S.~L.,  {Duncan} D.~K.,
  {Vaughan} A.~H.,  1984, \mn@doi [\apj] {10.1086/161945}, \href
  {http://adsabs.harvard.edu/abs/1984ApJ...279..763N} {279, 763}

\bibitem[\protect\citeauthoryear{{Pepe} et~al.,}{{Pepe} et~al.}{2002}]{Pepe02b}
{Pepe} F.,  et~al., 2002, The Messenger, \href
  {http://adsabs.harvard.edu/abs/2002Msngr.110....9P} {110, 9}

\bibitem[\protect\citeauthoryear{{Podolak} \& {Zucker}}{{Podolak} \&
  {Zucker}}{2004}]{Podo04}
{Podolak} M.,  {Zucker} S.,  2004, \mn@doi [Meteoritics and Planetary Science]
  {10.1111/j.1945-5100.2004.tb00081.x}, \href
  {http://adsabs.harvard.edu/abs/2004M%26PS...39.1859P} {39, 1859}

\bibitem[\protect\citeauthoryear{{Queloz} et~al.,}{{Queloz}
  et~al.}{2001}]{Que01}
{Queloz} D.,  et~al., 2001, \mn@doi [\aap] {10.1051/0004-6361:20011308}, \href
  {http://adsabs.harvard.edu/abs/2001A%26A...379..279Q} {379, 279}

\bibitem[\protect\citeauthoryear{{Queloz} et~al.,}{{Queloz}
  et~al.}{2009}]{Que09}
{Queloz} D.,  et~al., 2009, \mn@doi [\aap] {10.1051/0004-6361/200913096}, \href
  {http://adsabs.harvard.edu/abs/2009A%26A...506..303Q} {506, 303}

\bibitem[\protect\citeauthoryear{{Rajpaul}, {Aigrain}, {Osborne}, {Reece}  \&
  {Roberts}}{{Rajpaul} et~al.}{2015}]{Raj15}
{Rajpaul} V.,  {Aigrain} S.,  {Osborne} M.~A.,  {Reece} S.,   {Roberts} S.,
  2015, \mn@doi [\mnras] {10.1093/mnras/stv1428}, \href
  {http://adsabs.harvard.edu/abs/2015MNRAS.452.2269R} {452, 2269}

\bibitem[\protect\citeauthoryear{{Rajpaul}, {Aigrain}  \& {Roberts}}{{Rajpaul}
  et~al.}{2016}]{Raj16}
{Rajpaul} V.,  {Aigrain} S.,   {Roberts} S.,  2016, \mn@doi [\mnras]
  {10.1093/mnrasl/slv164}, \href
  {http://adsabs.harvard.edu/abs/2016MNRAS.456L...6R} {456, L6}

\bibitem[\protect\citeauthoryear{{Santerne} et~al.,}{{Santerne}
  et~al.}{2015}]{Sant15}
{Santerne} A.,  et~al., 2015, \mn@doi [\mnras] {10.1093/mnras/stv1080}, \href
  {http://adsabs.harvard.edu/abs/2015MNRAS.451.2337S} {451, 2337}

\bibitem[\protect\citeauthoryear{{Sasselov} \& {Lecar}}{{Sasselov} \&
  {Lecar}}{2000}]{Sas00}
{Sasselov} D.~D.,  {Lecar} M.,  2000, \mn@doi [\apj] {10.1086/308209}, \href
  {http://adsabs.harvard.edu/abs/2000ApJ...528..995S} {528, 995}

\bibitem[\protect\citeauthoryear{{Sato}, {Okuzumi}  \& {Ida}}{{Sato}
  et~al.}{2016}]{Sat16}
{Sato} T.,  {Okuzumi} S.,   {Ida} S.,  2016, \mn@doi [\aap]
  {10.1051/0004-6361/201527069}, \href
  {http://adsabs.harvard.edu/abs/2016A%26A...589A..15S} {589, A15}

\bibitem[\protect\citeauthoryear{{Schmitt} et~al.,}{{Schmitt}
  et~al.}{2014}]{Schm14}
{Schmitt} J.~R.,  et~al., 2014, \mn@doi [\apj] {10.1088/0004-637X/795/2/167},
  \href {http://adsabs.harvard.edu/abs/2014ApJ...795..167S} {795, 167}

\bibitem[\protect\citeauthoryear{{Schneider}, {Dedieu}, {Le Sidaner}, {Savalle}
   \& {Zolotukhin}}{{Schneider} et~al.}{2011}]{Sch11}
{Schneider} J.,  {Dedieu} C.,  {Le Sidaner} P.,  {Savalle} R.,   {Zolotukhin}
  I.,  2011, \mn@doi [\aap] {10.1051/0004-6361/201116713}, \href
  {http://adsabs.harvard.edu/abs/2011A\%26A...532A..79S} {532, A79}

\bibitem[\protect\citeauthoryear{{Seager}}{{Seager}}{2011}]{Sea11}
{Seager} S.,  2011, {Exoplanets}

\bibitem[\protect\citeauthoryear{{Sneden}}{{Sneden}}{1973}]{Sne73}
{Sneden} C.~A.,  1973, PhD thesis, The University of Texas at Austin.

\bibitem[\protect\citeauthoryear{{Sousa}}{{Sousa}}{2014}]{Sou14}
{Sousa} S.~G.,  2014, {ARES + MOOG: A Practical Overview of an Equivalent Width
  (EW) Method to Derive Stellar Parameters}.
pp 297--310, \mn@doi{10.1007/978-3-319-06956-2_26}

\bibitem[\protect\citeauthoryear{{Sousa}, {Santos}, {Israelian}, {Lovis},
  {Mayor}, {Silva}  \& {Udry}}{{Sousa} et~al.}{2011}]{Sou11a}
{Sousa} S.~G.,  {Santos} N.~C.,  {Israelian} G.,  {Lovis} C.,  {Mayor} M.,
  {Silva} P.~B.,   {Udry} S.,  2011, \mn@doi [\aap]
  {10.1051/0004-6361/201015646}, \href
  {http://adsabs.harvard.edu/abs/2011A\%26A...526A..99S} {526, A99+}

\bibitem[\protect\citeauthoryear{{Sousa} et~al.,}{{Sousa} et~al.}{2015}]{Sou15}
{Sousa} S.~G.,  et~al., 2015, \mn@doi [\aap] {10.1051/0004-6361/201425227},
  \href {http://adsabs.harvard.edu/abs/2015A%26A...576A..94S} {576, A94}

\bibitem[\protect\citeauthoryear{{Ter Braak}}{{Ter Braak}}{2006}]{TerB06}
{Ter Braak} C.~J.~F.,  2006, \mn@doi [Statistics and Computing, Volume 16,
  Issue 3, pp 239-249, 2006] {10.1007/s11222-006-8769-1}, \href
  {http://adsabs.harvard.edu/abs/2006S%26C....16..239T} {16}

\bibitem[\protect\citeauthoryear{{Thompson} et~al.,}{{Thompson}
  et~al.}{2018}]{Thom18}
{Thompson} S.~E.,  et~al., 2018, \mn@doi [\apjs] {10.3847/1538-4365/aab4f9},
  \href {http://adsabs.harvard.edu/abs/2018ApJS..235...38T} {235, 38}

\bibitem[\protect\citeauthoryear{{Torres}}{{Torres}}{2010}]{Tor10b}
{Torres} G.,  2010, \mn@doi [\aj] {10.1088/0004-6256/140/5/1158}, \href
  {http://adsabs.harvard.edu/abs/2010AJ....140.1158T} {140, 1158}

\bibitem[\protect\citeauthoryear{{Van Eylen}, {Agentoft}, {Lundkvist},
  {Kjeldsen}, {Owen}, {Fulton}, {Petigura}  \& {Snellen}}{{Van Eylen}
  et~al.}{2018}]{VanE18}
{Van Eylen} V.,  {Agentoft} C.,  {Lundkvist} M.~S.,  {Kjeldsen} H.,  {Owen}
  J.~E.,  {Fulton} B.~J.,  {Petigura} E.,   {Snellen} I.,  2018, \mn@doi
  [\mnras] {10.1093/mnras/sty1783}, \href
  {http://adsabs.harvard.edu/abs/2018MNRAS.tmp.1712V} {}

\bibitem[\protect\citeauthoryear{{Vanderburg} \& {Johnson}}{{Vanderburg} \&
  {Johnson}}{2014}]{Vand14}
{Vanderburg} A.,  {Johnson} J.~A.,  2014, \mn@doi [\pasp] {10.1086/678764},
  \href {http://adsabs.harvard.edu/abs/2014PASP..126..948V} {126, 948}

\bibitem[\protect\citeauthoryear{{Vanderburg} et~al.,}{{Vanderburg}
  et~al.}{2016a}]{Vand16a}
{Vanderburg} A.,  et~al., 2016a, \mn@doi [\apjs] {10.3847/0067-0049/222/1/14},
  \href {http://adsabs.harvard.edu/abs/2016ApJS..222...14V} {222, 14}

\bibitem[\protect\citeauthoryear{{Vanderburg} et~al.,}{{Vanderburg}
  et~al.}{2016b}]{Vand16b}
{Vanderburg} A.,  et~al., 2016b, \mn@doi [\apjl] {10.3847/2041-8205/827/1/L10},
  \href {http://adsabs.harvard.edu/abs/2016ApJ...827L..10V} {827, L10}

\bibitem[\protect\citeauthoryear{{Zeng} \& {Sasselov}}{{Zeng} \&
  {Sasselov}}{2013}]{Zeng13}
{Zeng} L.,  {Sasselov} D.,  2013, \mn@doi [\pasp] {10.1086/669163}, \href
  {http://adsabs.harvard.edu/abs/2013PASP..125..227Z} {125, 227}

\bibitem[\protect\citeauthoryear{{Zeng}, {Sasselov}  \& {Jacobsen}}{{Zeng}
  et~al.}{2016}]{Zeng16}
{Zeng} L.,  {Sasselov} D.~D.,   {Jacobsen} S.~B.,  2016, \mn@doi [\apj]
  {10.3847/0004-637X/819/2/127}, \href
  {http://adsabs.harvard.edu/abs/2016ApJ...819..127Z} {819, 127}

\bibitem[\protect\citeauthoryear{{Zeng}, {Jacobsen}  \& {Sasselov}}{{Zeng}
  et~al.}{2017a}]{Zeng17b}
{Zeng} L.,  {Jacobsen} S.~B.,   {Sasselov} D.~D.,  2017a, \mn@doi [Research
  Notes of the American Astronomical Society] {10.3847/2515-5172/aa9ed9}, \href
  {http://adsabs.harvard.edu/abs/2017RNAAS...1a..32Z} {1, 32}

\bibitem[\protect\citeauthoryear{{Zeng} et~al.,}{{Zeng}
  et~al.}{2017b}]{Zeng17a}
{Zeng} L.,  et~al., 2017b, in Lunar and Planetary Science Conference. p.~1576

\bibitem[\protect\citeauthoryear{{Zeng}, {Jacobsen}, {Sasselov}  \&
  {Vanderburg}}{{Zeng} et~al.}{2018}]{Zeng18}
{Zeng} L.,  {Jacobsen} S.~B.,  {Sasselov} D.~D.,   {Vanderburg} A.,  2018,
  \mn@doi [\mnras] {10.1093/mnras/sty1749}, \href
  {http://adsabs.harvard.edu/abs/2018MNRAS.tmp.1675Z} {}

\bibitem[\protect\citeauthoryear{{da Silva} et~al.,}{{da Silva}
  et~al.}{2006}]{Das06}
{da Silva} L.,  et~al., 2006, \mn@doi [\aap] {10.1051/0004-6361:20065105},
  \href {http://adsabs.harvard.edu/abs/2006A\%26A...458..609D} {458, 609}

\makeatother
\end{thebibliography}

%%%%%%%%%%%%%%%%%%%%%%%%%%%%%%%%%%%%%%%%%%%%%%%%%%

%%%%%%%%%%%%%%%%% APPENDICES %%%%%%%%%%%%%%%%%%%%%

%\appendix



%%%%%%%%%%%%%%%%%%%%%%%%%%%%%%%%%%%%%%%%%%%%%%%%%%

% Don't change these lines
\bsp	% typesetting comment
\label{lastpage}
\end{document}